\documentclass[conference]{IEEEtran}
\IEEEoverridecommandlockouts
\usepackage{cite}
\usepackage{amsmath,amssymb,amsfonts}
\usepackage{algorithmic}
\usepackage{graphicx}
\usepackage{textcomp}
\usepackage{xcolor}
\usepackage{booktabs}
\usepackage{float}
\usepackage{subcaption}

\def\BibTeX{{\rm B\kern-.05em{\sc i\kern-.025em b}\kern-.08em
    T\kern-.1667em\lower.7ex\hbox{E}\kern-.125emX}}
\begin{document}

\title{Graph Signal Adaptive Message Passing}
\author{
    \IEEEauthorblockN{
        Yi Yan, 
        Changran Peng, 
        Ercan E. Kuruoglu
    \thanks{This work is supported by Shenzhen Science and Technology Innovation Commission Grant JCYJ20220530143002005, Shenzhen Ubiquitous Data Enabling Key Lab Grant ZDSYS20220527171406015, and Tsinghua Shenzhen International Graduate School Start-up Fund QD2022024C. (Corresponding author: Ercan E. Kuruoglu.  E-mails: yiyiyiyan@outlook.com, kuruoglu@sz.tsinghua.edu.cn).}}
    \IEEEauthorblockA{
    Tsinghua-Berkeley Shenzhen Institute, Shenzhen International Graduate School, Tsinghua University}
}
\maketitle

\begin{abstract}
This paper proposes Graph Signal Adaptive Message Passing (GSAMP), a novel message passing method that simultaneously conducts online prediction, missing data imputation, and noise removal on time-varying graph signals. Unlike conventional Graph Signal Processing methods that apply the same filter to the entire graph, the spatiotemporal updates of GSAMP employ a distinct approach that utilizes localized computations at each node. This update is based on an adaptive solution obtained from an optimization problem designed to minimize the discrepancy between observed and estimated values. GSAMP effectively processes real-world, time-varying graph signals under Gaussian and impulsive noise conditions.
\end{abstract}
\begin{IEEEkeywords}
Graph Signal Processing, message passing, adaptive algorithms, Spatio-temporal estimation
\end{IEEEkeywords}
\section{Introduction}
Recently, the representation of irregularly structured multivariate signals as graphs has gained widespread popularity across diverse research fields due to the unparalleled ability of graphs to capture multivariate irregularities \cite{Ortega_graph_2018, Dong_Graph_ML_2020, Leus_2023_GSP}.
Time-varying graph signals are encountered in a broad spectrum in real life, with notable examples including monitoring the sea-surface temperature of the Pacific Ocean \cite{Castro_2023_time_varying_GNN}, observing fluctuations in stock market prices \cite{2024_Market_Qin}, recording wind speed at various locations \cite{Hong_GGARCH_2023}, analyzing traffic patterns within road networks \cite{Chen_2024_traffic}, examining states of brain activity across different regions \cite{zhao_2024_sequential}, and tracking the number of confirmed COVID-19 cases in different regions \cite{Giraldo_2022_time_varying_Sobolev}. 
Integrating classical adaptive filters with graph shift operations, along with smoothness and bandlimited filters, has been effective for online graph signal reconstruction by solving a convex optimization problem to minimize the error between observed and estimated signals.
Setting up the optimization problem using the $l_2$-norm leads to the Graph Least Mean Squares (GLMS) algorithm, which assumes Gaussian noise \cite{bib_LMS}.
However, the underlying noise in a variety of applications, including meteorological recordings \cite{1986_weather_impulsive} and powerline communication \cite{karakucs_2020_modelling}, is verified to possess impulsive behaviors, causing $ l_2$-norm algorithms to be unstable \cite{Chen_2016_variance}. 
As a result, the Graph-Sign (G-Sign) algorithm is proposed to use $l_1$-norm optimization instead of $l_2$-norm, providing robust estimations under impulsive noise \cite{yan_2022_sign}. 
The spatial diffusion versions of GLMS and G-Sign can be obtained by approximating the spectral filters using Chebyshev polynomials, resulting in the Graph Diffusion LMS (GDLMS) \cite{Roula_2017_LMS_Diffusion} and Graph-Sign-Diffusion (GSD) \cite{yan_2023_diffusion}.

The Graph Convolutional Neural Networks (GCN) is an extension of GSP on time-invariant tasks where nonlinearity is introduced to GSP by activation functions\cite{Dong_Graph_ML_2020, kipf2016semi, defferrard2016_cheb, bruna2013_spectral_GCN}.
GCN can be further generalized by the Message Passing Neural Networks (MPNN), where operations are defined locally on the node neighborhood relationships instead of globally on the entire graph \cite{gilmer_2017_neural, fey_2019_fast}. 
MPNN was proposed mostly as a performance-enhanced GCN or Graph Neural Network (GNN) in general for time-invariant tasks.
The Spatio-Temporal GCN (STGCN)  \cite{yu2018_STGCN} and several extensions such as the spectral-temporal GCN \cite{cao2020spectral} introduced GCN the ability to process time-varying data. 
However, GNNs, including STGCN and MPNNs, face several limitations compared to GSP methods, including lengthy training times, lower generalization capabilities, extensive data requirements for training, high resource intensity, implementation complexity, and reduced stability and robustness under noise. 
These drawbacks limit the practical deployment of GNNs in scenarios that demand straightforward, effective, and robust online estimation.
An adaptive GSP algorithm is needed for the online reconstruction of time-varying graph signals, benefiting from the low-cost, robust, and straightforward deployment of GSP alongside the representational flexibility of MPNNs.

With insights from the implementation simplicity of adaptive GSP algorithms and the expressiveness power of MPNNs, we take a step forward by breaking the convention of using only global information of the entire graph to define adaptive GSP methods.  
We propose the Graph Signal Adaptive Message Passing (GSAMP) algorithm, which is a novel adaptive graph algorithm defined using the localized node message passing scheme. 
Different from MPNNs which target time-invariant tasks and only updates based on the input data, GSAMP is obtained from a convex optimization problem based on the error between observation and estimation, leading to an adaptive update targeted towards time-varying tasks. 
The resulting GSAMP provides a comprehensive solution designed to address needs for online prediction, missing data imputation, and noise removal on time-varying graph signals. 
Additionally, under certain parameter settings, the GSAMP demonstrates high robustness under impulsive noise. 

\section{Preliminaries and Notation}
\label{sec_preliminaries}
A graph with $N$ nodes $\mathcal{V} = \{v_1 \dots v_N\}$ is denoted as $\mathcal{G}$. 
In this paper, for simplicity, we will consider the case of unweighted and undirected graphs. 
The time-varying function value defined on the nodes of $\mathcal{G}$ is a time-varying graph signal and is denoted as $\boldsymbol{x}[t]\in \mathbb{R}^N$ with $t$ being the time index.
The subscript of $x_v[t]$ is used to denote the signal on node $v$ at time $t$. 
The connection between nodes is represented by the adjacency matrix $\mathbf{A} \in \mathbb{R}^{N\times N}$ where the $v_i$ and $v_k$ is connected corresponds to the $ik^{th}$ entry being the 1.
The degree matrix $\mathbf{D} \in \mathbb{R}^{N\times N}$ is a diagonal matrix with each diagonal entry $d$ being the node degree calculated as the sum of the rows of $\mathbf{A}$. 
The graph Laplacian matrix is $\mathbf{L} = \mathbf{D} - \mathbf{A}$.
The Graph Fourier Transform (GFT) is the eigendecomposition of $\mathbf{L}$: $\mathbf{U \Lambda U}^T$, where $\mathbf{U}$ being the orthonormal eigenvectors and $\mathbf{\Lambda}$ being eigenvalues \cite{Ortega_2018}.
The eigenvalue-eigenvector pairs are sorted in increasing order, giving an analogy of frequency.
Filters in GSP define a function $h(\mathbf{\Lambda})$ in the spectral domain to manipulate frequency content through convolution $h(\mathbf{L})\boldsymbol{x}[t]=\mathbf{U} h(\mathbf{\Lambda}) \mathbf{U}^T \boldsymbol{x}[t]$, processing all node signals globally at the same time.
Missing node observations are modeled by a diagonal masking matrix $\mathbf{M}_\mathcal{S}$, where zeros on the diagonal indicate missing values and ones represent observed signals \cite{bib_LMS}.

The noise $\boldsymbol{\epsilon}[t]\in \mathbb{R}^N$ is modeled using the symmetric $\alpha$-stable (S$\alpha$S) distribution in this paper, where the three parameters characteristic exponent $\alpha$, location parameter $\mu$, and scale parameter $\gamma$ dictates the behavior of S$\alpha$S distribution \cite{Chen_2016_variance, herranz2004alpha, kuruoglu1997new}. 
Even though the S$\alpha$S has no analytic PDF, it has characteristic function $\boldsymbol{\phi}(t)=\exp\left\{j\eta t-\gamma|t|^\alpha\right\}$. 
The mean and variance of S$\alpha$S are undefined unless $1<\alpha\leqslant2$.
The S$\alpha$S becomes the Gaussian distribution when $\alpha$ = 2, making the mean being $\mu$ and variance being $2\gamma^2$.
The noisy partial observation of $\boldsymbol{x}[t]$ is denoted as $\boldsymbol{y}[t] = \mathbf{M}_\mathcal{S}(\boldsymbol{x}[t]+\boldsymbol{\epsilon}\left[t\right])$, the estimation of $\boldsymbol{x}[t]$ at time $t$ is denotes as $\hat{\boldsymbol{x}}[t]$.

\begin{figure}[htb]
    \centering
    \begin{subfigure}{.45\linewidth}
      \centering
    \includegraphics[trim= 270 375 250 375, width=\linewidth]{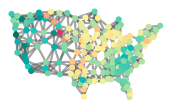}      
    \caption{Ground truth}
      \label{fig_top_1}
    \end{subfigure}%
    \begin{subfigure}{.45\linewidth}
      \centering
    \includegraphics[trim= 260 370 250 375, width=\linewidth]{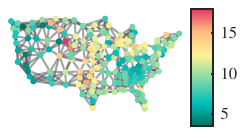}        
    \caption{GSAMP estimation}
      \label{fig_top_2}
    \end{subfigure}%
    \caption{A time instance of the time-varying wind speed.}
    \label{fig_top}
\end{figure}

\section{Methodology}
The message passing consists of the aggregated message $\boldsymbol{m}[t]$ and the update function $\Upsilon(.)$ \cite{gilmer_2017_neural}.
Here is a general formulation of the GSAMPS update function.
Each localized update at node $v$ in terms of $\hat{x}_v[t+1]$ and $m_v[t]$ is given by
\begin{equation}
    \hat{x}_v[t+1] = \Upsilon(x_v[t], m_v[t]) = \hat{x}_v[t] + m_v[t],
    \label{eq_update}
\end{equation}
where $m_v[t]$ is the localized aggregated message at node $v$ ($v^{th}$ element of $\boldsymbol{m}[t] \in \mathbb{R}^N$).
Assuming we input a graph signal $\boldsymbol{z}[t] \in \mathbb{R}^N$, all message passed to node $v$ in  $\mathcal{V}$ is denoted as
\begin{equation}
    m_v[t] = \Omega \left( \{ (z_v[t], z_j[t], w_{v,j}) \mid j \in \mathcal{N}_v \} \right),
    \label{eq_message}
\end{equation}
where the set of all the 1-hop neighboring nodes to $v$ is $\mathcal{N}_v$, the weight of the message between node $v$ and $j$ is $w_{v,j}$, and $\Omega(.)$ is the aggregation function.
Notice that there are several choices for the aggregation $\Omega(z_v[t], z_j[t], w_{v,j})$. 

\begin{figure}[tb]
    \centering
    \includegraphics[trim= 0 380 550 0, clip,width=0.8\linewidth]{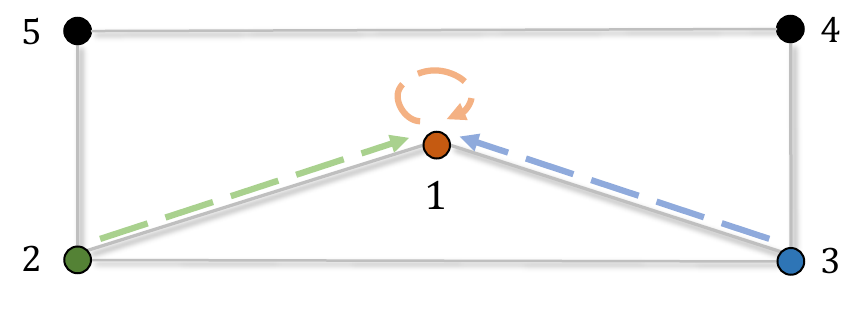}
    \caption{An illustration of localized message passing at node 1.}
    \label{fig_mp}
\end{figure}

The function $\Omega(.)$ is usually a permutation invariant operation suitable for aggregation on graphs with common choices found in previous literature including sum, mean, or max \cite{fey_2019_fast}. 
An illustration of node aggregation in message passing is shown in Fig.~\ref{fig_mp}.
For GASMP we consider the following three types of aggregations: weighted sum aggregation, median aggregation, and localized smoothness aggregation. 
The weighted sum aggregation is
\begin{equation}
    \Omega_{\text{sum}}(v) = \sum_{j \in \mathcal{N}_v} w_{v,j} z_j[t],
    \label{eq_message_sum}
\end{equation}
where $d_{vj}$ is the node degree of $v$.
Similarly, we can define the median aggregation as 
\begin{equation}
    \Omega_{\text{median}}(v) = \text{median} \left( \{z_j[t] \mid j \in \mathcal{N}_v\} \right).
    \label{eq_message_median}
\end{equation}
A localized smoothness aggregation applies the smoothness-based low pass filter $\mathbf{}$ to the subgraph $\mathcal{G}_v$ formed by node $v$ and its 1-hop neighborhood $\mathcal{N}_v$. 
We compose a vector $(\mathbf{z}_{\mathcal{N}_v}[t])$ as a vector that includes the signal on node $v$ and all signals from 1-hop neighboring nodes in $\mathcal{N}_v$.
We proceed by defining the graph Laplacian matrix of $\mathcal{G}_v$ as $\mathbf{L}_v$, then conduct the GFT $\mathbf{L}_v = \mathbf{U}_v\mathbf{\Lambda_v}\mathbf{U}_v^T$ to $\mathcal{G}_v$ to apply the low pass filter $h(\mathbf{\Lambda_v})$ by  $\mathbf{U}_vh(\mathbf{\Lambda_v})\mathbf{U}_v^T$. 
A low pass filter with passband $[0,\lambda_l]$ defines the filter $h(\mathbf{\Lambda}_v) = $diag$(\boldsymbol{h}_l)$ on the local neighborhood of node $v$, where the constant $l$ is the $l^{th}$ eigenvalue index in $\mathbf{\Lambda}_v$ and $\boldsymbol{h}_l$ is a $N$ by 1 vector with the first $l$ elements being 1 and last $N-l$ elements being 0. 
The resulting localized smoothness aggregation is denoted as
\begin{equation}
        \Omega_{\text{smooth}}(v) = \boldsymbol{\theta}_v \left( \boldsymbol{w}_v \odot \mathbf{z}_{\mathcal{N}_v}[t] \right)
,
        \label{eq_message_smooth}
\end{equation}
where $\odot$ is the element-wise multiplication and $\boldsymbol{\theta}_v$ corresponds to the row that node $v$ is in $\mathbf{U}_v h(\mathbf{\Lambda}_v) \mathbf{U}_v^T$.
The vector $\boldsymbol{w}_v$ is the weight vector of all the weights, including self-aggregation weights \(v\) and its $|\mathcal{N}_v|$ neighbors forming a vector \(\left[w_{v,v}, w_{v,1}, \ldots, w_{v,{|\mathcal{N}_v|}}\right]\).
In GSAMP, min or max aggregation is not considered as these definitions are primarily suited to categorical data and do not perform optimally with numerical graph signals.
It is worth mentioning that the mean aggregation can be achieved by setting the weights $w_{v,j}$ in the sum aggregation in \eqref{eq_message_sum} to $\frac{1}{d_{v}}$. 
Additionally, aggregated signals can then be combined linearly to form a more powerful representation, enhancing the ability of message passing to capture complex patterns in the graph signals \cite{gilmer_2017_neural}.

We would like the GSAMP update function in \eqref{eq_update} to follow an adaptive filter update logic.
To achieve online estimation, the message update should minimize the error between noisy partial observation $\boldsymbol{y}[t]$ and the estimation $\hat{\boldsymbol{x}}[t]$ by solving the following convex optimization problem:
\begin{equation}
        J\left(\hat{\boldsymbol{x}}[t]\right)=\frac{1}{p}\mathbb{E}\left\Vert \hat{\boldsymbol{x}}[t]-\boldsymbol{y}[t]\right\Vert_p^p.
        \label{eq_lmp_cost} 
\end{equation} 
Letting the message in \eqref{eq_message} be implemented from the solution of the cost function \eqref{eq_lmp_cost}, for $p = 2$, \eqref{eq_lmp_cost} becomes an $l_2$-norm optimization problem.
We can obtain an LMS version of GSAMP using stochastic gradient descent to solve \eqref{eq_lmp_cost}: 
\begin{equation}
    m_v^{(2)}[t] = \Omega \left( \{ (e^{(2)}_v[t], e^{(2)}_j[t], w_{v,j}) \mid j \in \mathcal{N}_v \} \right),
    \label{eq_lms_message}
\end{equation}
where $\boldsymbol{e}^{(2)}[t] =  \frac{\partial J(\hat{\boldsymbol{x}}[t])}{\partial \hat{\boldsymbol{x}}[t]} = \boldsymbol{x}[t]-\boldsymbol{y}[t]$, and the message function $\Omega()$ can be chosen from options such as \eqref{eq_message_sum}, \eqref{eq_message_median}, and \eqref{eq_message_smooth}.
Now, plugging \eqref{eq_lms_message} into \eqref{eq_update}, an adaptive message passing on graphs is formed similar to the classical adaptive LMS:
\begin{equation}
    \begin{split}
        \hat{\boldsymbol{x}}[t+1] = \left[\Upsilon(\hat{x}_1[t], m^{(2)}_1[t]) \dots \Upsilon(\hat{x}_N[t], m^{(2)}_N[t])\right]^T \\
        = \left[\hat{x}_1[t] + m^{(2)}_1[t] \dots \hat{x}_N[t] + m^{(2)}_N[t]\right]^T.
    \end{split}
        \label{eq_lms_update}
\end{equation}
We denote \eqref{eq_lms_update} as GSAMP-LMS, a message passing analogy of the classical adaptive LMS algorithm.

The $ l_2$-norm optimization implies the noise to be Gaussian but has been proven to be unstable under non-Gaussian impulsive noise, for example, the S$\alpha$S noise \cite{yan_2022_sign, Chen_2016_variance}. 
In order to form a robust message passing algorithm for time-varying graph signal reconstruction, we can set $p=1$ in \eqref{eq_lmp_cost} to form a $l_1$-norm optimization problem.
Then, the message becomes 
\begin{equation}
    m_v^{(1)}[t] = \Omega \left( \{ (e^{(1)}_v[t], e^{(1)}_j[t], w_{v,j}) \mid j \in \mathcal{N}_v \} \right),
    \label{eq_sign_message}
\end{equation}
where $\boldsymbol{e}^{(1)}[t] =  \text{sign}(\boldsymbol{x}[t]-\boldsymbol{y}[t])$.
Following the same steps of GSAMP-LMS derivations above, we can form the GSAMP-Sign algorithm by plugging \eqref{eq_sign_message} into \eqref{eq_update}:
\begin{equation}
    \begin{split}
        \hat{\boldsymbol{x}}[t+1] = \left[\Upsilon(\hat{x}_1[t], m^{(1)}_1[t]) \dots \Upsilon(\hat{x}_N[t], m^{(1)}_N[t])\right]^T \\
        = \left[\hat{x}_1[t] + m^{(1)}_1[t] \dots \hat{x}_N[t] + m^{(1)}_N[t]\right]^T.
    \end{split}
        \label{eq_sign_update}
\end{equation}

Comparing GSAMP-Sign with GSAMP-LMS, since the sign$()$ operation only has 3 possible outputs 1, 0, and -1, sign$()$ will clip the output of the message \cite{bib_sign_clip}. 
On one hand, this clipping remains unaffected by the impulsiveness of the noise due to its fixed magnitude, contributing to the robustness of the algorithm in impulsive noise environments. 
On the other hand, this same clipping introduces a trade-off between accuracy and robustness when the data is not subject to impulsive noise.
Additionally, the use of the sign$()$ and its output can reduce the runtime of the algorithms due to reduced computational complexity \cite{yan_2022_sign, bib_sign_clip}.
The choice between GSAMP-Sign and GSAMP-LMS is a design decision that depends on the specific requirements and conditions of the application.

In this paper, we define four types of weights for the values of $w_{v,j}$ between node \(v\) and its 1-hop neighboring node \(j\) based on their observation status:
\begin{enumerate}
    \item \(v\) and \(j\) are both observed (denoted as $w_{1}$),
    \item \(v\) is observed but \(j\) is missing (denoted as $w_{2}$),
    \item \(v\) is missing but \(j\) is observed (denoted as $w_{3}$),
    \item Both \(v\) and \(j\) are unobserved (denoted as $w_{4}$).
\end{enumerate}
The above weights allow GSAMP to weigh the contributions of messages from observed nodes and unobserved differently based on the local structures of each node. 
For the GSAMP to converge in a mean squared sense for steady-state estimations, the selection of weights $w$ should satisfy $\|\boldsymbol{m}[t]\|^2_2  \leq \|\boldsymbol{x}[t]\|^2_2$.  
One should bear in mind that the weights and aggregation can be different from what we discussed because message passing is a generalized formulation.
The formulation we provide for GSAMP in this paper aims for a straightforward realization of adaptive filters on graphs using GSAMP.

The message passing on node $v_i$ in the local neighborhood is naturally a data imputation algorithm because the missing data can be restored by properly setting the weights and then aggregating the neighborhood graph signal. 
Each GSAMP update \eqref{eq_lms_update} in or in \eqref{eq_sign_update} is a message passing based on the estimation error controlled by the weights $\boldsymbol{w}$ in the direction opposite to the difference between $\hat{\boldsymbol{x}}[t]$ and $\boldsymbol{y}[t]$. 
Even though we defined the message passing in GSAMP using only 1-hop aggregations, repeating the aggregations process $k$ times essentially is a $k$-hop message passing \cite{ZHOU202057}.  

In conventional GSP the operations are global and symmetric because $\mathbf{U} h(\mathbf{\Lambda}) \mathbf{U}^T$ is symmetric and is applied to the entire graph signal. 
However, notice that in GSAMP the definition of $w_{v, j}$ can be set asymmetrically, for example, when $w_2 \neq w_3$. 
GSAMP can realize adaptive GSP algorithms, as it represents a more generalized formulation. 
For example, GSAMP reduces to GLMS when all message weights are set to the same value $w$ and $\boldsymbol{m}[t] = w \mathbf{U} h(\mathbf{\Lambda}) \mathbf{U} \hat{\boldsymbol{x}}[t]$, which eliminates localized computation and weight flexibility. 
The key distinction is that GSAMP allows individually tuned weights $\boldsymbol{w}$ and the freedom of defining various types of aggregations for each node, while adaptive GSP algorithms reply on globally defined filters.
In other words, the localized weights, messages, and aggregations make GSAMP fundamentally different from adaptive GSP algorithms that update the entire graph uniformly. 
What distinguishes GSAMP from classical adaptive filters is that the weights of classical LMS are directly learned from the signal, but in GSAMP the weights are associated with the topological structure of the local neighborhood of each node.

\begin{table*}[htb]
    \centering
    \begin{tabular}{c|ccccccc}
    \toprule
       S$\alpha$S Noise Parameters  & GSAMP (sum) & GSAMP (median) & GSAMP (smooth) & GLMS & G-Sign & GDLMS & GSD\\
         \midrule
   $\alpha = 2, \gamma = 0.1, \mu = 0$ & \underline{307} & 341 & \textbf{305} & 325 & 356 & 315 & 335  \\ 
    $\alpha = 2, \gamma = 0.15, \mu = 0$ & \underline{309} & 344 & \textbf{306 }& 329 & 359 & 316 & 336  \\
    $\alpha = 2, \gamma = 0.2, \mu = 0$ & \underline{313} & 348 & \textbf{308} & 335 & 364 & 318 & 337    \\
             \midrule

       $\alpha = 1.3, \gamma = 0.1, \mu = 0$ & \textbf{345} & 440 & \underline{423} & 3850 & 584 & 776 & 505  \\ 
    $\alpha = 1.3, \gamma = 0.15, \mu = 0$ & \textbf{356} & \underline{491} & 579 & 8255 & 890 & 1353 & 755  \\
    $\alpha = 1.3, \gamma = 0.2, \mu = 0$ & \textbf{376} & \underline{582} & 813 & 14422 & 1324 & 2161 & 1125    \\
         \bottomrule
    \end{tabular}
    \caption{Average of the MSE over all time points for all tested algorithms with the best-performing algorithms in bold and second-best-performing underlined. Note that for settings $\alpha = 2$ GSAMP realization is the GSAMP-LMS in \eqref{eq_lms_update}; for settings $\alpha = 1.3$ GSAMP realization is the GSAMP-Sign in \eqref{eq_sign_update}.}
    \label{table_MSE}
\end{table*}

\section{Experimental results and discussion}
\label{experiments}
The capability of GSAMP to conduct online estimation is tested by giving only the noisy observation containing missing values at time $t$ and letting GSAMP form the graph signal at time $t+1$.
The dataset in the experiment is a time-varying graph signal consisting of hourly wind speed records from 197 weather stations across the United States \cite{bib_dataset}.
The graph topology is constructed using the method seen in \cite{Spelta_2020_NLMS} where the topology is formed by nearest neighbors based on locations of the station.
We configured the observation mask $\mathbf{M}_\mathcal{S}$ as a time-invariant mask, maintaining 130 out of the 197 nodes as observed using the greedy method that maximizes the spectral graph signal content shown in \cite{Spelta_2020_NLMS}. 
In this configuration, the task for the tested algorithm is to perform both temporal predictions and spatial data imputation, since missing observations are imputed using the graph topological information.
Figure~\ref{fig_top_1} depicts a single time instance of the dataset.

\begin{figure}[h]
  \centering
    \includegraphics[trim = {180 310 180 310}, clip, width=\linewidth]{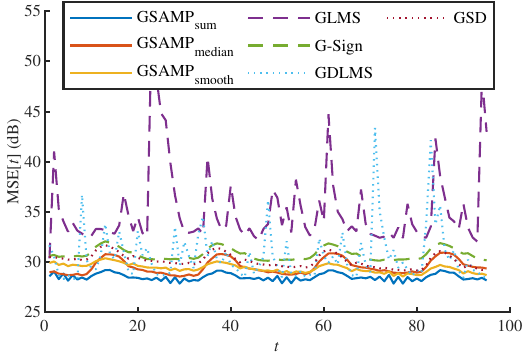}
          \vspace{-15 pt}
    \caption{The MSE from $t = 1$ to $95$ $\alpha = 1.3$ and $ \gamma = 0.1$.}
    \label{fig_MSE}
\end{figure}

The noise is added S$\alpha$S noise with six different configurations: for $\alpha = 1.3$ and $2$ we tested one of each with $\gamma = 0.1, 0.15,$ and $0.2$, all with $\mu = 0$.
We apply GSAMP-LMS in \eqref{eq_lms_update} to the cases of $\alpha = 2$ and GSAMP-Sign in \eqref{eq_sign_update} to the cases of $\alpha = 1.3$.
There are three message configurations of GSAMP: weighed sum in \eqref{eq_message_sum}, median in \eqref{eq_message_median}, and localized smoothness in \eqref{eq_message_smooth}.
The weights \((w_1, w_2, w_3, w_4)\) of GSAMP configurations for sum, median, and smooth aggregations are \((1, 0, 2, 0)\), \((0.7, 0, 0.7, 0)\), and \((0.7, 0, 1.95, 0)\), respectively. 
The GSAMP is compared against GLMS \cite{bib_LMS}, GDLMS \cite{Roula_2017_LMS_Diffusion}, G-Sign \cite{yan_2022_sign}, and GSD \cite{yan_2023_diffusion}; the step size parameter of GLMS, GDLMS, and GSD are $1.6$; for G-Sign the step size is $1.3$. 
The parameter selection is done by grid search by applying the algorithms to the first 1/3 time points on the data.
For the algorithms that require a filter, we adopted an ideal low-pass filter with the cutoff frequency of $0.4\lambda_N$.
All tested algorithms follow a diffusion initialization shown in \cite{yan_2023_diffusion} where the missing nodes are filled in with a weighted averaging of the neighborhood observed signal.
The performance of all algorithms is measured using the mean squared error (MSE) at each time step between the estimation value and the ground truth graph signal: MSE$[t] = \frac1N\sum_{i=1}^N{(x_i[t]-\hat{x}_i[t])^2}$. 
Here, subscript $i$ indicates the $i^{th}$ node in the graph.
The MSEs of all the algorithms are calculated at each time point for the forecasted temperature as illustrated in Fig.~\ref{fig_MSE} for the setting $\alpha = 1.3$ and $ \gamma = 0.1$. 
The average of the MSE$[t]$ results throughout all 95 time points are shown in Table~\ref{table_MSE}. 
The estimation using GSAMP-LMS to the instance shown in Fig.~\ref{fig_top_1} is shown in Fig.~\ref{fig_top_2} for the case $\alpha = 2$ and $\gamma = 0.1$.
The algorithms are repeated 100 times in MATLAB 2024a.

Analyzing Table~\ref{table_MSE}, GSAMP-LMS in \eqref{eq_lms_update} combined with localized smoothness aggregation in \eqref{eq_message_smooth} achieves the lowest MSE for the cases $\alpha = 2$.
GSAMP-Sign in \eqref{eq_sign_update} combined with weighted sum aggregation in \eqref{eq_message_sum} achieves the lowest MSE for the cases $\alpha = 1.3$.
The overall higher MSE of adaptive GSP algorithms can be attributed to their global definition of the filters. 
In GSAMP, each message has distinct values derived from calculating the error, which is the solution from the optimization problem \eqref{eq_lmp_cost} similar to adaptive GSP algorithms.
Message passing not only enhances representation flexibility but also increases expressiveness in utilizing localized information from each graph node.
Each of the messages is also assigned different weights $w$, granting GSAMP additional degrees of freedom that are absent in traditional adaptive GSP algorithms.
Another interesting observation from Fig.~\ref{fig_MSE} is that when the S$\alpha$S parameters are impulsive ($\alpha$ = 1.3) rather than Gaussian ($\alpha = 2$), the $l_1$-norm algorithms are more stable than $l_2$-norm algorithms, which is as expected, reflecting the fact that $l_1$-norm optimization is more suitable than $l_2$-norm optimization under impulsive noise. 

\section{Conclusion and Future Work}
The GSAMP is proposed for time-varying graph signal estimation by adopting various types of localized node message passing schemes. 
The two realizations of GSAMP, namely GSAMP-LMS and GSAMP-Sign, have been confirmed to effectively leverage message passing and adaptive filters to create localized, flexible, and expressive algorithms for processing time-varying graph signals.
In addition, $l_1$-norm optimization enhances the robustness of GSAMP-Sign at handling impulsive noise scenarios.
Future directions include exploring the combination of GSAMP with learning algorithms to learn the weights and to combine federated learning for a distributed implementation of GSAMP. 
Future applications of GSAMP could extend to fields such as material science \cite{SAFFARIMIANDOAB2021115197} and image separation \cite{costagli2007image}, which are important in advancing artificial intelligence and scientific research.
\nopagebreak

\pagebreak
\bibliographystyle{IEEEbib}
\bibliography{refs.bib}
\end{document}